\documentstyle[aps,preprint]{revtex}
\begin{document}

\draft
\tightenlines
\makebox[\textwidth][r]{SNUTP 97-108}

\begin{center}
{\large\bf
Zero Temperature Chiral Phase Transition in
(2+1)-Dimensional QED with a Chern-Simons Term}
\vskip 0.2in
Deog Ki Hong\footnote{Email address:$\>$\tt dkhong@hyowon.cc.pusan.ac.kr}\\
\vskip 0.2in
\it
Department of Physics, Pusan National
University \protect\\
Pusan 609-735, Korea
\end{center}

\begin{abstract}
We investigate the zero temperature chiral phase transition
in (2+1)-dimensional QED in the presence of a Chern-Simons term,
changing the number of fermion flavors.
In the symmetric phase, there are no light degrees of freedom
even at the critical point. Unlike the case without a Chern-Simons
term, the phase transition is first-order.

\end{abstract}


Recently, the phase transition at zero temperature  
has been studied extensively in both supersymmetric~\cite{sw} 
and non-supersymmetric gauge 
theories~\cite{twa95,twa96,miransky97,chivukula97}.
Gauge theories show a rich phase structure  
as one changes the number of color $N_C$  
or the number of flavors $N_f$ in different representations.   
In a non-supersymmetric $SU(N_C)$ gauge theory, 
the phase structure is infered 
from the behavior of gauge coupling for different number of
flavors.  For large $N_f$, the theory is in Coulomb phase and 
for $N_f$ just below $11N_C/2$ it has an infrared fixed  
point~\cite{banks}. As $N_f$ is reduced further, chiral symmetry 
breaking occurs. When $N_f$ crosses the critical point $N_f^c$
for chiral symmetry breaking, the theory exhibits  
discontinuity in the mass of scalar particles~\cite{twa96}.
This discontinuity is then further investigated in view of conformal 
symmetry~\cite{miransky97}. 
In this brief report, we study the zero temperature phase 
transition associated with dynamical mass generation  
in (2+1)-dimensional QED with a Chern-Simons term and will show 
that the mass spectrum in scalar particles does change
discontinuously near the critical point, which  
therefore admits the conformal symmetry argument 
advocated recently by Miransky and Yamawaki~\cite{miransky97}.  

Quantum electrodynamics in three dimensions describes   
the high temperature limit of (3+1)-dimensional QED and 
also certain planar condensed matter systems.
In studying those systems, it is important to understand 
how fermions get mass dynamically. 
Dirac mass term for two-component complex fermion in 2+1 dimensions
breaks parity ($P$) and time reversal symmetry ($T$), as well as 
the continuous flavor symmetry of the system. 
Pisarski found, solving the Schwinger-Dyson (SD) gap equation in 
$1/N_f$ expansion, that dynamical mass generation occurs 
and the flavor symmetry $U(2N_f)$ breaks down to 
$U(N_f)\times U(N_f)$~\cite{pisarski}. Appelquist {\it et al.} 
then analyzed the SD equation more carefully to find that 
there is a phase transition at a critical number of fermion flavors, 
below which dynamical mass generation takes place~\cite{twa88}.
In the four-component spinor notation, in the leading 
order in $1/N_f$ expansion, it is $32/\pi^2$ in Landau gauge. 
Later, the exact value was found to be $128/(3\pi^2)$~\cite{nash}.    
Similar analysis has been done for (2+1)-dimensional QED 
with a Chern-Simons term,  
which allows photon to have a gauge-invariant, but $P$ and 
$T$ violating mass~\cite{jackiw}, and it is found that 
the Chern-Simons term reduces the critical number of flavor and the 
magnitude of dynamical fermion mass, namely the critical 
number of flavor becomes $\tilde N_f^c=N_f^c/
\left[1+(\kappa/\alpha)^2\right]$ and the dynamical mass 
$m(\kappa\ne0)=m(\kappa=0)\exp\left(-4N_f/N_f^c
\cdot\kappa^2/\alpha^2\right)$~\cite{my93}. 
It is then further investigated by Kondo and Maris~\cite{kondo95}
to show that the dynamical mass does not vanish as $N_f$  
approaches the critical value, just like the order parameter 
of first-order phase transition.  

As was done by Appelquist {\it et al.}~\cite{twa95,twa96},  
we follow the method employed by Nambu and Jona-Lasinio~\cite{nambu}
to solve the SD equation for the fermion-antifermion 
scattering amplitude in the symmetric phase 
of (2+1)-dimensional QED with a Chern-Simons term with $N$ 
two-component complex fermions in $1/N$ expansion. 
To facilitate $1/N$ expansion, we keep $e^2N\equiv16\alpha$ finite 
when $N$ goes to infinity.   
We set the (Euclidean) momentum of the initial fermion and 
antifermion to $q/2$, but assign momenta $q/2\pm p$ for the 
final fermion and antifermion, allowing a momentum transfer $p$ 
between the fermion and antifermion. We then look for a pole 
in (Minkowsky) $q^2$ for the scattering amplitude. 

We take the Dirac indices of the initial fermion and antifermion 
as $\lambda$ and $\rho$, while the final state fermion and antifermion 
$\sigma$ and $\tau$. In the leading order in $1/N$,
the SD equation for the fermion-antifermion 
scattering amplitude is given in Euclidean space as   
\begin{eqnarray}
&{}&T_{\lambda\rho\sigma\tau}(p,q)  = {16\alpha\over N} 
\left(\gamma^{\mu}\right)_{\sigma\lambda}D_{\mu\nu}(p)
\left(\gamma^{\nu}\right)_{\rho\tau} \nonumber\\
&{}&\quad\quad+ {16\alpha\over N}\int_k T_{\lambda\rho{\sigma^{\prime}}
{\tau^{\prime}}}(k,q)\left(\gamma^{\mu}
{1\over {1\over2}\mathord{\not\mathrel{q}}+
\mathord{\not\mathrel{k}}}\right)_{\sigma\sigma^{\prime}}
D_{\mu\nu}(p-k)\left(
{1\over -{1\over2}\mathord{\not\mathrel{q}}+
\mathord{\not\mathrel{k}}}\gamma^{\nu}\right)_{\tau^{\prime}\tau}, 
\label{sd}
\end{eqnarray}
where $D_{\mu\nu}(p)$ is the photon propagator and 
Dirac gamma matrices $\gamma^{\mu}$ in three dimensions 
are just the
Pauli matrices  satisfying ${\rm
tr}(\gamma^{\mu}\gamma^{\nu}\gamma^{\lambda})
=2i\epsilon^{\mu\nu\lambda}$.  The photon propagator in $1/N$
expansion is given by summing up  all the bubble diagrams as 
\begin{equation}
D_{\mu\nu}=\left[g_{\mu\nu}-p_{\mu}p_{\nu}/p^2\right]\Pi_1(p)+
\epsilon_{\mu\nu\lambda}p^{\lambda}\Pi_2(p),
\label{propagator}
\end{equation}
where we choose Landau gauge and 
\begin{eqnarray}
\Pi_1(p)&=&{\left|p\right|+\alpha\over\left|p\right|
\left[(\left|p\right|+\alpha)^2+{\kappa}^2\right]} 
\nonumber\\  
\Pi_2(p)&=&{1\over p^2}\cdot 
{\kappa\over (\left|p\right|+\alpha)^2+{\kappa}^2}
\end{eqnarray} 
with $\kappa$ the 
coefficient of the Chern-Simons term.  
(Note that in general there will be a one-loop correction 
to the Chern-Simons term but we have chosen a regulator such that 
the correction will be of the order of $1/N$.) 
Since the direct product of two spinors in 2+1 dimensions 
is either scalar or vector, we may write
the scattering amplitude as 
\begin{equation}
T_{\lambda\rho\sigma\tau}=\delta_{\lambda\rho}\delta_{\sigma\tau}T+
\delta_{\lambda\rho}(\gamma^{\alpha})_{\sigma\tau}T_{\alpha}+
(\gamma^{\alpha})_{\lambda\rho}\delta_{\sigma\tau}T_{\alpha}^{\prime}
+(\gamma^{\alpha})_{\lambda\rho}(\gamma^{\beta})_{\sigma\tau}
T_{\alpha\beta},
\label{fullamp}
\end{equation}
where $T$ is the scalar channel amplitude, $T_{\alpha}$ vector
channel, and so on. Further, we write the vector channel amplitude as 
\begin{equation}
T^{\alpha}(p,q)=ip^{\alpha}T_1(p,q)+iq^{\alpha}T_1^{\prime}(p,q).
\label{vectoramp}
\end{equation} 
For small $q$ limit, the
second term in Eq.~(\ref{vectoramp}) is negligible . 

Plugging Eq.~(\ref{fullamp}) into the SD equation,
one finds the scalar channel amplitude $T$ is coupled to 
the vector channel amplitude $T^{\alpha}$. 
We consider $p\gg q$. Then $q$ will be simply act as an infrared 
cutoff in the loop integrals. 
Multyplying $\delta_{\lambda\rho}\delta_{\sigma\tau}/4$ and  
$\delta_{\lambda\rho}\gamma^{\alpha}_{\tau\sigma}/4$ respectively to 
Eq.~(\ref{sd}), and 
integrating the angular variables, we get for $p\ll\alpha$ 
\begin{eqnarray}
T(p,q)&=&{16\over N_0p}+{8\over N_0\pi^2}\int_q^{\infty}{dk\over pk}
T(k,q)\left(p+k-\left|p-k\right|\right)\nonumber\\
&{}&-{4\over N_1\pi^2}\int_q^{\infty}dk\left[{p^2-k^2\over
2pk}\ln\left({p+k\over\left|p-k\right|}\right)\right]T_1(k,q)
\label{scalar}\\
T_1(p,q)&=&-{16\over N_1 p^2}-{8\over N_1\pi^2p^2}\int_q^{\infty}dk
\left[{p^2-k^2\over
2pk}\ln\left({p+k\over\left|p-k\right|}\right)\right]T(k,q)
\\
\label{vector}
\end{eqnarray}
where $p$ and $q$ denote the magnitude of momentum and 
\begin{equation}
N_0=N\cdot {\alpha^2+{\kappa}^2\over \alpha^2}
=N_1\cdot{\kappa\over\alpha}.
\end{equation}

When the Chern-Simons term is absent, (2+1)-dimensional QED is finite 
in the ultraviolet (UV) region and 
the integral in Eq. (\ref{sd}) will be rapidly damped for
$k>\alpha$. With a Chern-Simons term, the UV structure
of (2+1)-dimensional QED may be quite different. Especially, in the
broken phase, the Chern-Simons term will dominate in the
ultraviolet region, since the fermion mass function falls off
slowly~\cite{kondo95}. But, in the symmetric phase, where we are
interested in, the parity-even mass function is zero and in the
photon propagator, Eq. (\ref{propagator}), the parity-odd
part falls off more rapidly than the parity-even part.
Therefore, the integral in Eq. (\ref{sd})  
falls off rapidly for $k>\alpha,\kappa$.
Therefore, we can take $\alpha$ as the UV cutoff 
and assume $\kappa\simeq\alpha$. 
As approximation, we further take 
\begin{equation}
{p^2-k^2\over 2pk}\ln\left({p+k\over\left|p-k\right|}\right)
\simeq \theta(p-k)-\theta(k-p),
\end{equation}
which should give a good approximation since 
the error in the integration will be the order of $p/\alpha$. 

Now, we can convert the coupled integral equations, 
Eq.'s~(\ref{scalar}) and (\ref{vector}),   
into coupled differential equations: 
\begin{eqnarray}
 {d^2\over dp^2}\left(pT\right) & = & -{16\over N_0\pi^2p}T
-{8\over N_1\pi^2p}{d\over dp}\left(p^2T_1\right) \label{scalard}\\
{d\over dp}\left(p^2T_1\right) & = &-{16\over N_1\pi^2}T.  
\label{vectord}
\end{eqnarray}
We see that the vector channel amplitude is subdominant by $1/N_1$, 
compared to the scalar channel amplitude,  
and so is the effect of the Chern-Simons term to the scalar 
amplitude. In $1/N$ expansion,
we can neglect the second term in the right-hand side 
of Eq.~(\ref{scalard}) and  
the scalar channel amplitude is decoupled;
\begin{equation}
 {d^2\over dp^2}\left(pT\right) = -{16\over N_0\pi^2p}T,
\end{equation}
which admits a power solution,
\begin{equation}
T(p,q)={A(q)\over\alpha}\left({p\over\alpha}\right)^{-(1/2)+(1/2)\eta}
+{B(q)\over \alpha}\left({p\over\alpha}\right)^{-(1/2)-(1/2)\eta},
\label{sols}
\end{equation} 
where $\eta=\sqrt{1-\tilde N_f^c/N}$ and 
$\tilde N_f^c={64\over\pi^2}/ 
\left[1+(\kappa/\alpha)^2\right]$. (Note in the four-component 
spinor notation $\tilde N_f^c$ has to be reduced by 1/2.) 
With the solution for $T$, we find from Eq.~(\ref{vectord})
\begin{equation}
T_1(p,q)={C_1(q)\over\alpha^2}\left({p\over\alpha}\right)^{-2}
+{A_1(q)\over\alpha^2}\left({p\over\alpha}\right)^{-(3/2)+(1/2)\eta}
+{B_1(q)\over\alpha^2}\left({p\over\alpha}\right)^{-(3/2)-(1/2)\eta},
\label{solv} 
\end{equation}
where
\begin{eqnarray}
 A_1(q) & = & -{32\over N_1\pi^2}{A(q)\over 1+\eta},\quad 
B_1(q)  =  -{32\over N_1\pi^2}{B(q)\over 1-\eta} \\
C_1(q) & = & -{16\over N_1}+{16\over N_1\pi^2}\left\{
{A(q)\over 1+\eta}\left[\left({q\over\alpha}\right)^{(1/2)+(1/2)\eta}
+1\right]+{B(q)\over 1-\eta}
\left[\left({q\over\alpha}\right)^{(1/2)-(1/2)\eta}+1\right]
\right\}.
\end{eqnarray}
The $q$ dependence of the amplitudes can be determined by substituting 
the solutions back into Eq.~(\ref{scalar}). This gives  
\begin{eqnarray}
 A(q) & = & {(1+\eta)\pi^2/2(q/\alpha)^{-(1/2)+(1/2)\eta}
\over (q/\alpha)^{\eta}- [(1+\eta)/(1-\eta)]^2}\\
B(q) & = & -{1+\eta\over1-\eta}A(q).
\end{eqnarray}
We see that the pole structure of the scalar and vector  
channel amplitudes in (2+1)-dimensional QED is same whether or not  
the Chern-Simons term is present, except that the exponent $\eta$   
changes due to the change in the critical number of fermion flavors.  
The location of the poles of the amplitudes in the complex 
$q$ plane is at $q=q_0$ with 
\begin{equation}
 \left|q_0\right|=\alpha\left({1+\eta\over1-\eta}\right)^{2/\eta}. 
\end{equation}
When $N$ approaches the critical number ($N\to\tilde N_f^c$), 
$\eta\to0$ and 
$\left|q_0\right|\to\alpha\exp(4)$, 
which is beyond the domain of our approximation, 
$q\ll\alpha$. The mass of the scalar and vector bound states
does not vanish as $N$ approaches the critical value, while in the 
broken phase the scalar bound state mass is zero due to 
Goldstone theorem. We find therefore that, 
with or without a Chern-Simons term, (2+1)-dimensional QED exhibits
discontinuity in the scalar mass spectrum. This finding shows that 
there is no long range correlation in symmetric phase, because  
both photons and fermions also get parity-odd mass  
due to the Chern-Simons term. 
Since the order parameter in the broken phase for the 
zero temperature chiral phase transition does not vanish 
even at the critical point, we see that the phase transition 
associated with dynamical mass generation behaves as 
a regular first-order phase transition.  
This is in  sharp contrast to pure QED3, 
which exhibits a peculiar phase transition; 
namely, in pure QED3, the order parameter in the broken phase 
vanishes at the critical point while there is no scalar 
whose physical mass approaches zero at the critical point 
in the symmetric phase.  
In Ref.~\cite{twa95}, Appelquist {\it et al.} argued that 
the peculiarity of the phase transition can be attributed to the 
fact that the effective potential is not analytic at zero dynamical 
mass, $m$, of the fermion at zero momentum because the theory 
has a long range force mediated by gauge fields. Here, we note that 
the regular first-order phase transition in QED3
with a Chern-Simons term 
supports this argument because the effective potential is analytic
at $m=0$ when a Chern-Simons term is present; 
there is only a finite number of oscillating solutions 
to the SD gap equation  with the Chern-Simons 
term~\cite{kondo95,Ebihara} and there is no long range 
force.  

To conclude, we have studied the zero temperature chiral phase 
transition in QED3 with a Chern-Simons term by solving a 
Schwinger-Dyson equation for the fermion-antifermion 
scattering amplitude in the symmetric phase ($N>\tilde N_f^c$) in 
$1/N$ expansion. 
We find that as in pure QED3 no light degrees of freedom appear in 
the symmetric phase, which implies that the phase transition is  
first-order because the order-parameter does not vanish  
in the broken phase even at the critical point. Since 
the effective potential is analytic in the dynamical fermion mass 
at zero momentum, this result supports 
the argument of Appelquist {\it et al.} that the peculiar phase 
transiton at zero temperature in pure QED3 is 
due to the non-analyticity of the effective potential 
at zero dynamical fermion mass.

\acknowledgments

We thank T. Appelquist for useful discussions.
This work was supported in part by the KOSEF through SRC program of 
SNU-CTP and also by Basic Science Research Program, Ministry of 
Education, 1996 (BSRI-96-2413).

\end{document}